\documentclass{svproc}

\usepackage{url}

\usepackage[table,dvipsnames]{xcolor}
\usepackage{graphicx}
\usepackage[utf8]{inputenc}
\usepackage[T1]{fontenc}

\usepackage{subeqnarray}
\usepackage{amsmath}
\usepackage{lineno,lipsum}
\usepackage{hyperref}
\usepackage{cleveref}
\crefname{equation}{Eq.}{Eqs.}
\crefname{section}{Sect.}{Sects.}
\crefname{figure}{Fig.}{Figs.}

\graphicspath{{./figures/}{./}}

\makeatletter
\newcommand\footnoteref[1]{\protected@xdef\@thefnmark{\ref{#1}}\@footnotemark}
\makeatother

\usepackage{amsfonts}

\newcommand{\be}{\begin{equation}}
\newcommand{\ee}{\end{equation}} 
\newcommand{\bea}{\begin{eqnarray}}
\newcommand{\eea}{\end{eqnarray}}
\newcommand{\pois}{\textrm{Pois}}

\newcommand{\f}[2]{\frac{#1}{#2}}
\newcommand{\bup}[1]{\left(#1\right)}
\newcommand{\rup}[1]{\left[#1\right]}
\newcommand{\Exp}{\mathbb{E}}
\newcommand{\Ag}{\mathbf{A^\mathrm{(g)}}}
\newcommand{\Ao}{\mathbf{A^\mathrm{(o)}}}
\newcommand{\Agij}{A^\mathrm{(g)}_{ij}}
\newcommand{\Agji}{A^\mathrm{(g)}_{ji}}
\newcommand{\Aoij}{A^\mathrm{(o)}_{ij}}

\newcommand{\Z}{\mathbf{Z}}
\newcommand{\Zij}{Z_{ij}}
\newcommand{\lij}{\lambda_{ij}}
\newcommand{\Qij}{Q_{ij}}
\newcommand{\elbo}{\mathcal{L}}

\newcommand{\xp}{\mbox{{\small \textsc{EXP}}}}
\newcommand{\nxp}{\mbox{{\small \textsc{NoEXP}}}}
\newcommand{\mt}{\mbox{{\small \textsc{MultiTensor} }}}

\begin{document}
\mainmatter              % start of a contribution
\title{Modeling Node Exposure for Community Detection in Networks
}
\titlerunning{Node Exposure}  % abbreviated title (for running head)
%                                     also used for the TOC unless
%                                     \toctitle is used
%
\author{Sameh Othman\inst{1} \and Johannes Schulz\inst{1}
 \and Marco Baity-Jesi\inst{2} \and Caterina De Bacco\inst{1}}
\authorrunning{Sameh Othman et al.} % abbreviated author list (for running head)
%
%%%% list of authors for the TOC (use if author list has to be modified)
\tocauthor{Sameh Othman, Johannes Schulz, Marco Baity-Jesi and Caterina De Bacco}
\institute{Max Planck Institute for Intelligent Systems, Cyber Valley, Tuebingen, 72076, Germany
\and Eawag (ETH), \"Uberlandstrasse 133, CH-8600 D\"ubendorf, Switzerland}

\maketitle              % typeset the title of the contribution

\begin{abstract}

In community detection, datasets often suffer a sampling bias for which nodes which would normally have a high affinity appear to have zero affinity. This happens for example when two affine users of a social network were not exposed to one another. Community detection on this kind of data suffers then from considering affine nodes as not affine.
To solve this problem, we explicitly model the (non-)exposure mechanism in a Bayesian community detection framework, by introducing a set of additional hidden variables. 
Compared to approaches which do not model exposure, our method is able to better reconstruct the input graph, while maintaining a similar performance in recovering communities. Importantly, it allows to estimate the probability that two nodes have been exposed, a possibility not available with standard models.
\end{abstract}

\section{Introduction}
Modeling the mechanisms of how nodes interact in networks is a relevant problem in many applications. In social networks, we observe a set of interactions between people, and one can use this information to cluster them into communities based on some notion of similarity~\cite{fortunato2010community}. Broadly speaking, the connections between users can be used to infer users' membership, and this in turns determines the likelihood that a pair of users interacts. Real networks are often sparse, people interact with a tiny amount of individuals, compared to the large set of possible interactions that they could in principle explore.
Traditionally, models for community detection in networks treat an existing link as a positive endorsement between individuals: if two people are friends in a social network, this means they like each other. In assortative communities, where similar nodes are more likely to be in the same group \cite{newman2018networks,fortunato2016community}, this encourages the algorithm to put these two nodes into the same community. On the contrary, a non-existing link influences the model to place them into different communities, as if the two non-interacting individuals were not compatible.  
However, many of these non-existing
links (--especially in large-scale networks--) are absent because the individuals are not aware of each other, rather than because they
are not interested in interacting. This is a general problem in many network data sets: we know that interacting nodes have a high affinity, but we can not conclude the contrary about non-interacting nodes.\\
This problem has been explored in the context of recommender systems \cite{liang2016modeling,yang2018unbiased,wang2016learning,chuklin2015click}, where it is crucial to learn what items that a user did not consume could be of interest. In this context, items' exposure is often modeled by means of propensity scores or selection biases assigned to user-item pairs that increase the probability of rare consumption events. \\ It is not clear how to adapt these techniques to the case of networks of interacting individuals, hence the investigation of this problem in the context of networks is still missing. Existing approaches partially account for this by giving more weight to existing links, as in probabilistic generative models that use a Poisson distribution for modeling the network adjacency matrix~\cite{de2017community,ball2011efficient,zhao2012consistency,schein2016bayesian}. These methods are effective, but may be missing important information contained in non-existing links.

\section{Community detection with exposure}
We address this problem by considering a probabilistic formulation that assigns probabilities to pairs of nodes of being exposed or not. These are then integrated into standard probabilistic approaches for generative networks with communities. For this, as a reference model we consider \mt\ \cite{de2017community}, as it is a flexible model that takes in input a variety of network structures (\textit{e.g.} directed or undirected networks, weighted or unweighted) and detects overlapping communities in a principled and scalable way.

\subsection{Representing exposure}\label{sec:rep-exp}

Consider an $N\times N$ \emph{observed} network adjacency matrix $\Ao$, where $\Aoij\geq0$ is the weight of the interaction between nodes $i$ and $j$, this is the input data. For instance, $\Aoij$ could be the number of times that $i$ and $j$ met or exchanged messages. 
If a link $\Aoij$ exists, this indicates an affinity between individuals $i$ and $j$, triggered by both individuals' inner preferences. If the link does not exist ($\Aoij=0$), one usually assumes that this indicates a lack of affinity between $i$ and $j$. However, the link might not exist simply because $i$ and $j$ never met.
This is the case in social networks, where an \textit{ego} might follow an \textit{alter} because of personal preference, but this choice is subject to being exposed to the \textit{alter} in the first place. This suggests that the event of being exposed to someone influences the patterns of interactions observed in networks. 
We are interested in incorporating this notion of exposure in modeling network data, and investigate how results change.

\begin{figure}[t!]
\centering
 \includegraphics[width=\textwidth]{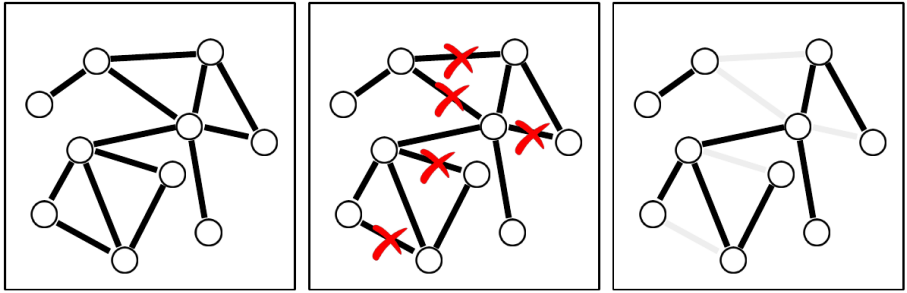}
 \caption{Diagram of the exposure mechanism. On the \textbf{left} we have the full graph ($\Ag$). We then set to zero the probability of some connections through a mask $\Z$ (\textbf{center}), and reconstruct the true graph and communities based solely on the visible links, $\Ao$ (\textbf{right}).}
 \label{fig:diagram}
\end{figure}

To represent this, we postulate the existence of a \emph{ground-truth adjacency matrix}, $\Ag$, that indicates the affinity between nodes $i$ and $j$ regardless of whether the two nodes were exposed to each other (Fig.~\ref{fig:diagram}--left). In addition, we introduce a \emph{dilution matrix} $\Z$ (red crosses in Fig.~\ref{fig:diagram}--center), with values $\Zij=0,1$ indicating whether nodes $i$ and $j$ were exposed ($\Zij=1$) or not ($\Zij=0$).
The observed matrix is then the element-wise product of the ground truth network times the dilution matrix,
\be
\Ao = \Ag \otimes \Z\,,
\ee
where $\otimes$ indicates an element-by-element multiplication. A diagram of the resulting matrix is shown in Fig.~\ref{fig:diagram}--right.
Through this representation, a zero-entry $\Aoij=0$ can be attributed to $\Agij=0$ (lack of affinity), $\Zij=0$ (lack of exposure) or both.

Standard models for community detection do not account for exposure, therefore they treat a zero-entry $\Aoij=0$ as a signal for non-affinity. 
We aim at measuring both communities and exposure, given the observed data $\Aoij$. In other words, for a given node $i$, we would like to estimate its community membership and for a given pair $(i,j)$ we want to estimate the probability that they were exposed to each other. 
For simplicity, we show derivations for the case of undirected networks, but similar ones apply to directed ones.

\subsection{The ground truth adjacency matrix}
In our notation, we use $\theta$ to denote the latent variables affecting community detection, \textit{i.e.} determining the probability of observing an interaction between $i$ and $j$ given that they have been exposed. 
We will treat the case of symmetric edges, $\Agij=\Agji$, and provide an extension to asymmetric interactions in App.~\ref{app:asym}.
Following the formalism of Ref.~\cite{de2017community}, we assign a $K$-dimensional hidden variable $u_i$ to every node $i$. Since different communities may interact in different ways, we also introduce a $K\times K$ affinity matrix $w$, regulating the density of interactions between different groups. The latent variables related to the ground truth matrix are then $\theta=(u,w)$. 

We express the expected interaction between two nodes through a parameter 
\begin{equation}\label{eq:lambda}
\lij=\sum_{k,q}^K u_{ik}u_{jq} w_{kq}\,,    
\end{equation}
and extract the elements of $\Ag$ from a Poisson distribution with mean $\lij$,
\be
P(\Agij|u_{i},u_{j},w)=\pois \bup{\Agij;\lij} = \frac{e^{-\lij}\lij^{\Agij}}{\Agij!}\,.
\ee
We then assume conditional independence between different pairs of edges given the latent variables $P(\Ag|u,u,w)=\prod_{i<j}P(\Agij|u_{i},u_{j},w)$, but this can be generalized to more complex dependencies \cite{safdari2021generative,contisciani2021community,safdari2022reciprocity}. We do not explore this here.
\subsection{The observed adjacency matrix}
The observed adjacency matrix depends on whether two nodes were exposed or not, through the matrix $\Z$. 
If $\Zij=1$, the  two nodes are exposed, and the edge comes from the ground truth matrix, \textit{i.e.} $P(\Aoij| \Zij=1, \theta) = P(\Agij|\theta) = \pois(\Agij; \lij)$.
If $\Zij=0$, then $\Aoij=0$ regardless of $\lij$.
Therefore, the elements of $\Ao$ are extracted from the distribution
\be\label{eqn:likelihood}
P(\Aoij| \Zij, \theta) =  \pois(\Aoij; \lij)^{\Zij}\, \delta(\Aoij)^{1-\Zij} \,.
\ee

Since $\Zij$ is binary, we assign it a Bernoulli prior with parameter $\mu_{ij}$,
\bea\label{eqn:PZ}
P(\Z|\mu) = 
\prod_{i<j}P(\Zij|\mu_{ij}) = \prod_{i<j}\bup{\mu_{ij}}^{\Zij}\, \bup{1-\mu_{ij}}^{1-\Zij}\,.
\eea
The parameter $\mu_{ij}$ will depend on some latent variable related to nodes $i$ and $j$. There are several possible choices for that. Here, we consider a simple setting:
\begin{align} \label{eqn:prior}
\mu_{ij}&=\mu_i\,\mu_j\,,\\
\mu_{i}&\in[0,1]\,,
\end{align}
This allows to keep the number of parameters small and has an easy interpretation.
In fact, the parameter $\mu_{i}$ acts as the propensity of an individual to be exposed to others: the higher its value, the higher the probability that node $i$ will be exposed to other nodes. This way of modeling exposure only adds one more parameter per node, allowing for heterogeneous behaviors among users while keeping the model compressed.
The full set of variables that need to be inferred consists of the $u$, the $w$ and the $\mu$ variables, which amounts to $NK+K^2+N$ parameters, which is one order of magnitude smaller than the $ N^2$ elements of $\Ao$.

\subsection{Inference and Expectation-Maximization}
Given the data $\Ao$, our goal is to first determine the values of the parameters $\theta$, which  fixes the relationship between the hidden indicator $\Zij$ and the data, and then to approximate $\Zij$ given the estimated $\theta$.

We perform this using statistical inference as follows. Consider the posterior distribution $P(\Z,\theta| \Ao)$. Since the dilution $\Z$ is independent from the parameters $\theta$ and all the edges are considered conditionally independent given the parameters, Bayes' formula gives
\be\label{eqn:joint}
P(\Z,\theta| \Ao) = \f{P(\Ao|\Z,\theta) P(\Z |\mu) P(\theta)}{P(\Ao)} \,.
\ee
Summing over all the possible indicators we have:
\be
P(\theta| \Ao) = \sum_{\Z}P(\Z,\theta| \Ao) = \prod_{i<j}^N\sum_{\Zij=0,1} P(\Zij,\theta| \Ao) \quad,
\ee
which is the quantity that we need to maximize to extract the optimal $\theta$.
It is more convenient to maximize its logarithm, as the two maxima coincide. We use Jensen's inequality:
\be\label{eqn:jensen}
\log P(\theta| \Ao) = \log\sum_{\Z}P(\Z,\theta| \Ao) \geq  \sum_{\Z} q(\Z)\, \log \f{P(\Z,\theta| \Ao)}{q(\Z)}:= \mathcal{L}(q,\theta,\mu)\,
\ee
where $q(\Z)$ is {any distribution satisfying $\sum_{\Z} q(\Z)=1$, we refer to this as the variational distribution}.

Inequality~(\ref{eqn:jensen}) is saturated when
\be\label{eqn:q}
q(\Z) = \f{P(\Z,\theta| \Ao)}{\displaystyle\sum_{\Z} P(\Z,\theta| \Ao)}\,,
\ee
hence this choice of $q$ maximizes $\mathcal{L}(q,\theta,\mu)$ with respect to $q$.
Further maximizing it with respect to $\theta$ gives us the optimal latent variables. This can be done in an iterative way using Expectation-Maximization (EM), alternating between maximizing with respect to $q$ using \Cref{eqn:q} and then maximizing $\mathcal{L}(q,\theta,\mu)$ with respect to $\theta$ and $\mu$.

To obtain the updates for the parameters we need to derive the equations that maximize $\mathcal{L}(q,\theta,\mu)$ with respect to $\theta$ and $\mu$ and set these derivatives to zero. This leads to the following closed-form updates:
\begin{align}
u_{ik} &= \f{\sum_{j} Q_{ij}\, A_{ij}\sum_{q}\rho_{ijkq} }{\sum_{j}Q_{ij}\, \sum_{q}u_{jq}w_{kq}} \label{eqn:u} \\
w_{kq} & = \f{\sum_{i,j} Q_{ij}\, A_{ij}\rho_{ijkq} }{\sum_{i,j}Q_{ij}\, u_{ik}u_{jq}} \label{eqn:w}  \\
\rho_{ijkq} & =\f{u_{ik}u_{jq}w_{kq}}{\sum_{k,q}u_{ik}u_{jq}w_{kq}} \label{eqn:rho} \\
\mu_{i} & =\f{\sum_{j}Q_{ij}} {\sum_{j}\f{(1-Q_{ij})\, \mu_{j} }{ (1-\mu_{i}\, \mu_{j})}} \label{eqn:mu} \quad,
\end{align}
where we defined $\Qij=\sum_{\Z}\,\Zij\,q(\Z) $ the expected value of $Z_{ij}$ over the variational distribution. 

As $\mu_i$ appears on both sides of \Cref{eqn:mu}, this  can be solved with root-finding methods bounding $\mu_i$ to the interval $[0,1]$, to be compatible as a parameter of the Bernoulli prior.\footnote{
In practice, we limit the domain of $\mu_i$ to the interval [$\epsilon,1-\epsilon]$, where $\epsilon$ is a small hyperparameter chosen to avoid numerical overflows of $\elbo$. To maintain the model interpretable in terms of exposure, at the end of the optimization we set to zero each $\mu_{i}\equiv \epsilon$ and to one each $\mu_{i} \equiv 1-\epsilon$.
}

\begin{figure}\centering
 \includegraphics[width=0.325\textwidth]{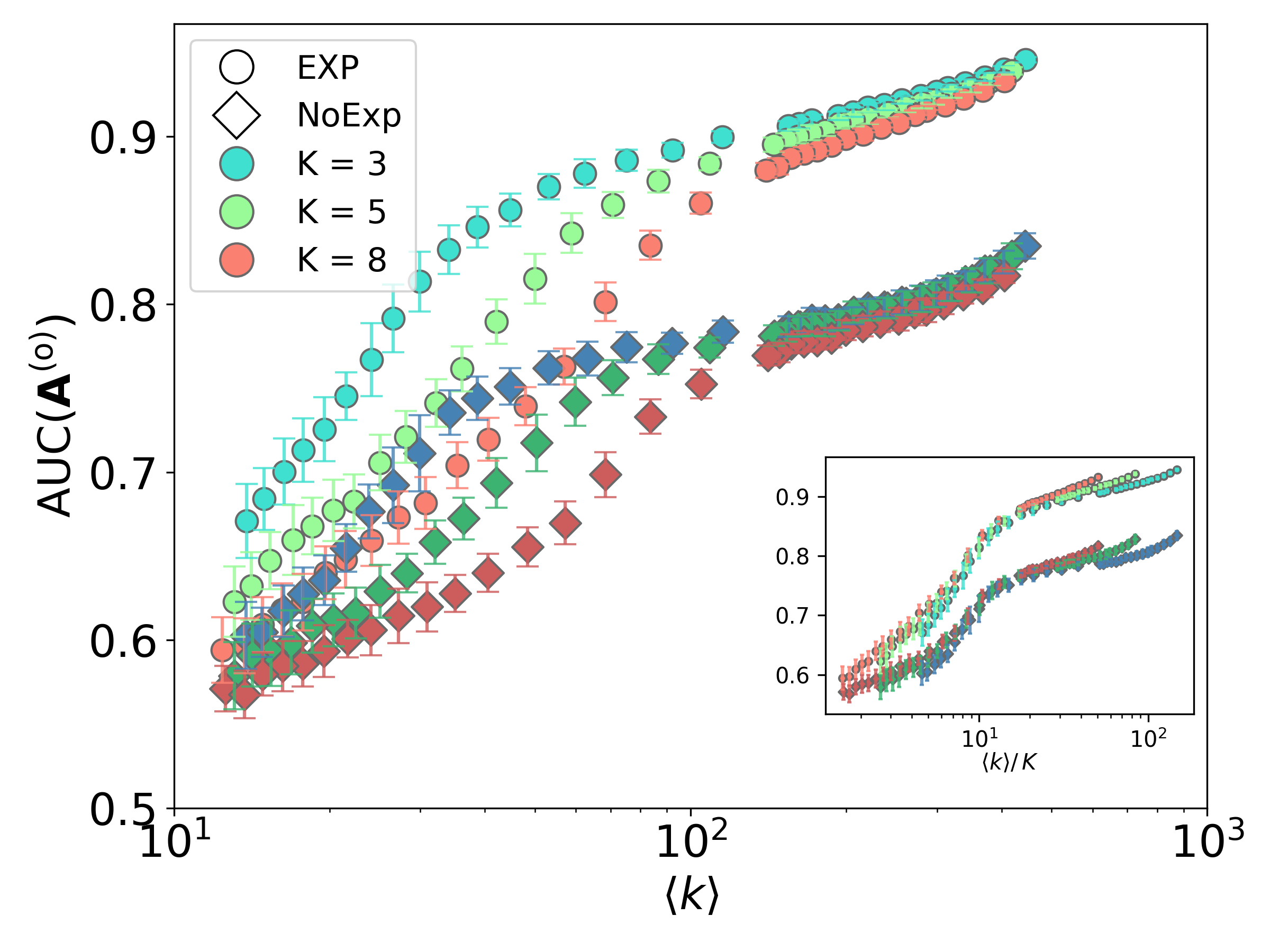}
 \includegraphics[width=0.325\textwidth]{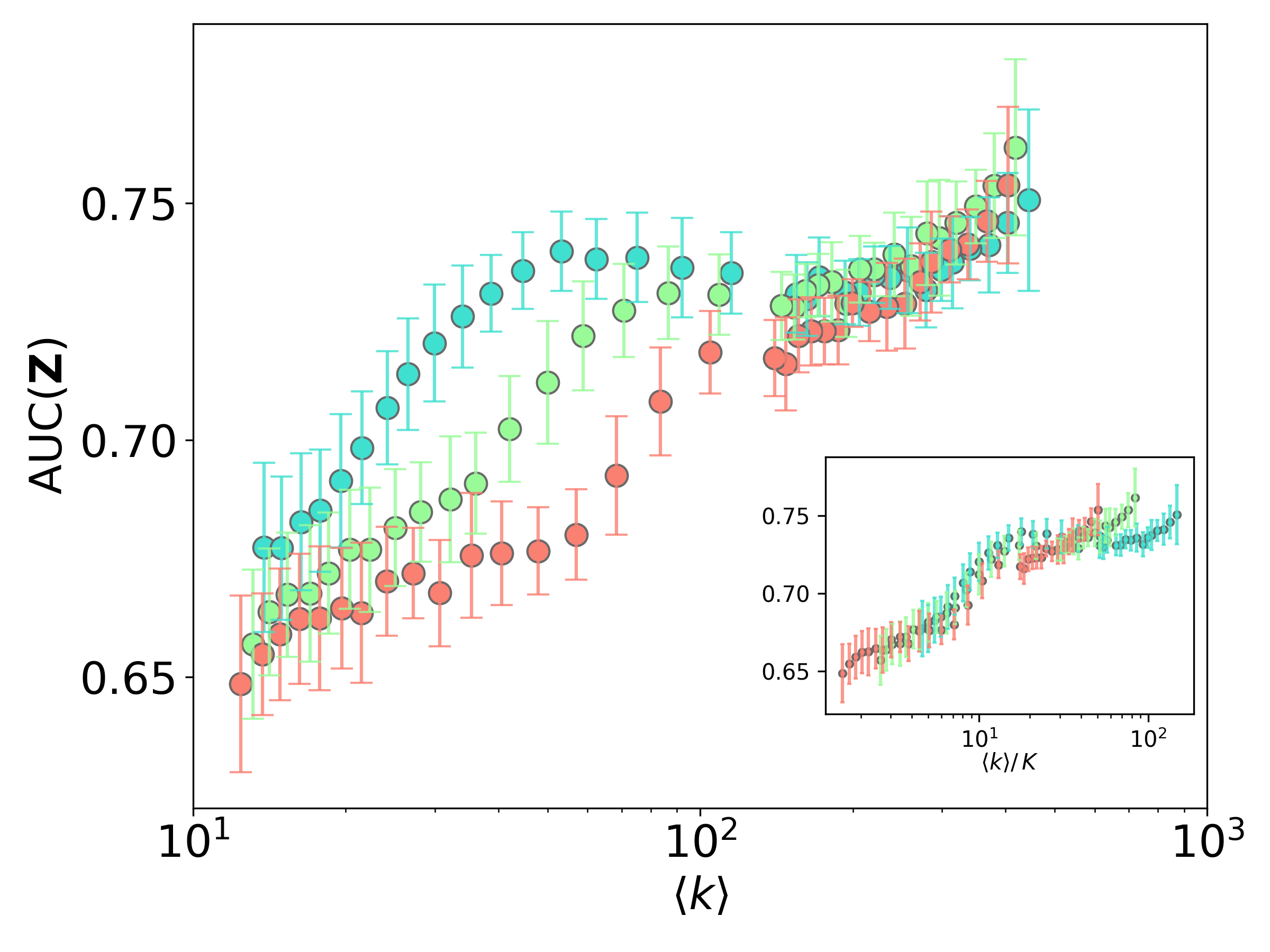}
 \includegraphics[width=0.325\textwidth]{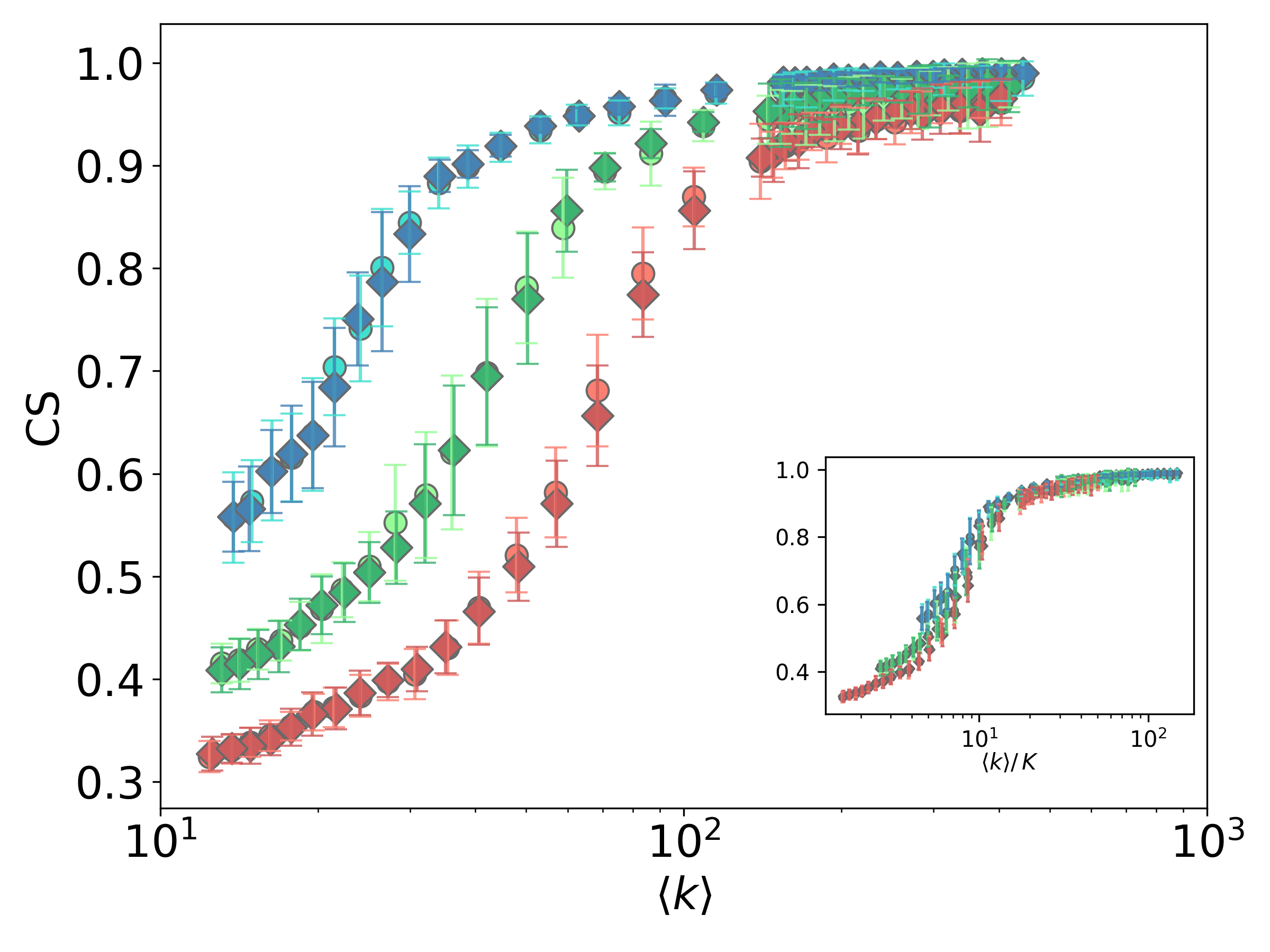}
 \caption{
 Performance of the \xp\ and \nxp\ algorithms on synthetic data. The matrix $\Ag$ has $K=3,5,8$ communities and $N=500$. The exposure mask $Z$ is extracted from a binomial distribution with parameter $\mu_{ij}=\mu_i\mu_j$.
 \textbf{Left:} AUC between the inferred values and the ground truth used to generate $\Ao$.
 \textbf{Center:} AUC of the reconstruction of the exposure mask $\Z$.
 \textbf{Right:} Cosine similarity between inferred and ground truth communities.
 \textbf{Inset:} We show the same data as in the main plots by rescaling the average number of links by the number of communities.
 }
 \label{fig:synth1}
\end{figure}

Finally, to evaluate $q(Z)$, we substitute the estimated parameters inside \Cref{eqn:joint}, and then into \Cref{eqn:q} to obtain:
\begin{align}
q(\Z)&={\prod_{i<j}}\, \Qij^{\Zij}\, (1-\Qij)^{(1-\Zij)}\,\label{eqn:qij},
\end{align}
where
\be\label{eqn:Qij}
\Qij = \f{\pois(A_{ij};\lambda_{ij})\mu_{ij}}{\pois(A_{ij};\lambda_{ij})\mu_{ij}+\delta(A_{ij})\, (1-\mu_{ij})} \quad.
\ee
In other words, the optimal {$q(\Z)$ is a product $\prod_{i<j}q_{ij}(\Z_{ij})$ of } Bernoulli distributions $q_{ij}$ with parameters $\Qij$. {This parameter is also a point-estimate of the exposure variable, as for the Bernoulli distribution $\Qij = \Exp_{q}\rup{\Z_{ij}}$.\\
{The algorithmic EM procedure then works by initializing at random all the parameters and then iterating \Crefrange{eqn:u}{eqn:mu} for fixed $q$, and the calculating \Cref{eqn:Qij} given the other parameters, and so on until convergence of $\elbo$.  The function $\elbo$ is not convex, hence we are not guaranteed to converge to the global optimum. In practice, one needs to run the algorithm several times with different random initial parameters' configurations and then select the run that leads to best values of $\elbo$. In the following experiments we use 5 of such realizations.}

\section{Results}\label{sec:res}
We test our algorithm on synthetic and real data, and compare it to its formulation without exposure, \textit{i.e.} the \mt algorithm described in Ref.~\cite{de2017community}.
In the following, we refer to our algorithm  as \xp, and we use \nxp\ for the algorithm that does not utilize exposure.

\subsection{Synthetic data}\label{sec:synt}
Synthetic data experiments are particularly interesting, because we can validate our model performances on the ground truth values.
The creation of a synthetic dataset follows the generative model described in \Cref{sec:rep-exp}:
\begin{enumerate}
\item  For a graph with $N=500$ nodes, we generate the latent parameters $\theta$ and $\mu$ as follows. We draw overlapping communities by sampling $u_i$ from a Dirichlet distribution with parameter $\alpha_k=1$, $\forall k$;
we choose an assortative $w$ by selecting the off-diagonal entries to be 0.001 times smaller then the on-diagonal ones. We then vary $K\in [3,5,8]$. 
We draw $\mu_i$ from a Beta distribution $\mathrm{Beta}(\mu_i;2,\beta)$, where we vary $\beta\in [0.1,10]$ to tune the fraction of unexposed links.
 \item Sample $\Ag_{ij}$ from a Poisson distribution with means $\lambda_{ij}=\sum_{k,q}u_{ik}u_{jq}w_{kq}$.
 \item Sample $\Z$ from a Bernoulli distribution of means $\mu_{ij}=\mu_{i}\mu_j$.
 \item Calculate the matrix $\Ao=\Ag\otimes\Z$. This matrix has on average $\langle k\rangle$ links per node.
\end{enumerate}
We repeat this procedure 10 times for each set of parameters to obtain different random realizations of synthetic data. We then apply the \xp\ and \nxp\ algorithms to $\Ao$ to learn the parameters and study the performance as a function of $\langle k\rangle$, controlling the density of observed edges. 

\paragraph{Reconstructing hidden links}
We start by testing the ability of the model to predict missing links, a procedure often used as a powerful evaluation framework for comparing different models \cite{liben2007link,lu2011link}. 
We use a 5-fold cross-validation scheme where we hide $20\%$ of the edges in $\Ao$ and train the model on the remaining $80\%$. Performance is then computed on the hidden $20\%$ of the edges.
As a performance evaluation metric we measure the area under the receiver operating characteristic curve (AUC) between the inferred values and the ground truth used to generate $\Ao$ on the test set. The AUC is the probability that a randomly selected existing edge is predicted with a higher score than a randomly selected non-existing edge. A value of 1 means optimal performance, while 0.5 is equivalent to random guessing. As the score of an edge $\Ao_{i,j}$ we use the quantity $Q_{ij}\, \lambda_{ij}$ for \xp, and $\lambda_{ij}$ for \nxp. In both cases, these are the expected values of $\Aoij$ using the estimates of the latent parameters and, for \xp, over the inferred $q(\Z)$. We find that the \xp\ algorithm outperforms \nxp\ by a large margin, which increases as the network becomes more dense, going above 10\%, as shown in \Cref{fig:synth1}--left. At low densities, the performance increase of the \xp\ algorithm is narrow for models with a large number of communities, while at large densities it becomes bigger and independent of the number of communities. This result suggests that \xp\ is capturing the input data better--consistently for varying dilution densities--than a model that does not account for exposure.}

\paragraph{Guessing unexposed links}
Our algorithm not only allows us to predict missing edges but also gives interpretable estimates of the probability of exposure between nodes. These probabilities follow naturally from the posterior distribution on \textbf{Z}, which is the Bernoulli distribution in \Cref{eqn:qij}. Standard algorithms as \nxp\ cannot estimate this. We can use the mean value $Q_{ij}$ as in \Cref{eqn:Qij} as a score of an edge to compute the AUC between inferred and ground truth values of $\Z$, analogously to what was done for reconstructing $\Ao$. 
We report in \Cref{fig:synth1}--center the ability of \xp\ to reconstruct the matrix $\Z$, \textit{i.e.} to infer which edges were removed in the dilution step. The AUC varies between $0.65$ and $0.75$, well above the random baseline of $0.5$. We notice how the values increase as the density of connection increases, but stay above $0.65$ even at small density values, where reconstruction is more challenging.

\paragraph{Inferring communities}
In \Cref{fig:synth1}--right, we can see that \xp\ and \nxp\ show similar performances in reconstructing communities.
From this plot we can also notice how reconstruction improves for larger densities and fewer communities. The similar performances may be due to selecting a simple prior as in \Cref{eqn:prior}. For a more structured prior, the inferred communities would likely change and potentially improve. Given this similar community detection abilities but the better predictive power in reconstructing $\Ao$, we argue that the learned $Q_{ij}$'s are important to boost prediction compared to a model that does not properly account for exposure. This is true even for a simple prior.

\paragraph{Dependence on the number of communities}
All of these metrics exhibit a scaling w.r.t. the variable  $\langle k\rangle/K$, as can be seen in the insets of \Cref{fig:synth1}. This suggests that the curves seem to be independent of the number of communities when accounting for this rescaling. Thus observing the behavior for one particular value of $K$ should be informative enough to understand how the model behaves for various densities.

\begin{figure}[t!]
\centering
 \includegraphics[width=0.7\textwidth]{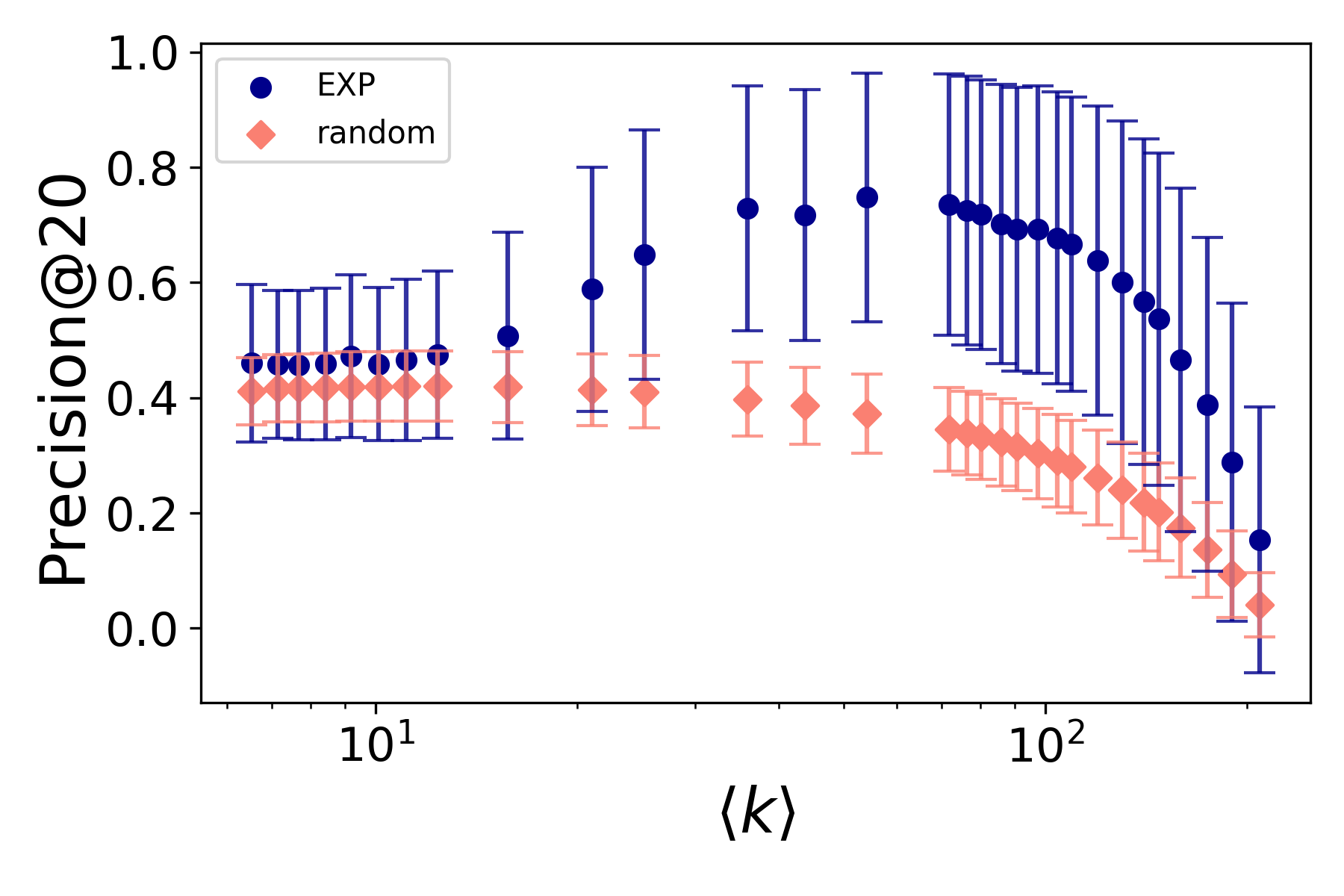}
 \caption{Suggesting unexposed compatible nodes. For each node $i$, we suggest the 20 links with highest $\lij$ inferred by our algorithm from the non-observed links where $\Aoij=0$. We show the P@20 averaged across all nodes and compare with a uniform-at-random baseline (random) where 20 nodes are selected at random among the available ones. Error bars are standard deviations. Here we use a synthetic network generated as in \Cref{sec:synt} with $N=500$ and $K=5$.} 
 \label{fig:precision}
\end{figure}

\paragraph{Suggesting good matches}
Since the \xp\ algorithm is good at predicting which nodes were removed from the original graph (\Cref{fig:synth1}--center), we can use this to address the following question: Is the \xp\ algorithm able to suggest two nodes that have high affinity despite not having any connection? In other words, we are asking whether we are able to find links that in $\Ao$ are absent, but have a high expected value in $\Ag$.
To test this ability, we take for each node $i$: a) all the possible neighbors $j$ such that $\Aoij=0$; b) select among them the 20 with the largest inferred affinity $\lij$; and c) check how many of those are present in $\Ag$. We call Precision@20 (P@20) the fraction of links which were correctly inferred, averaging across all nodes. In \Cref{fig:precision} we show that for intermediate dilution values, the P@20 reaches around 80\%, and outperforms random guessing at any value of the dilution. Notice that random guessing in not constant in $\langle k \rangle$. This is because this depends on the number of missing links in $\Ao$, and those depend both on the density of $\Ag$ and on the dilution mask $\Z$. Specifically, P@20 of the random baseline goes as $(\langle k \rangle_g-\langle k \rangle )/(N-\langle k \rangle)$, where $\langle k \rangle_g$ is degree of $\Ag$. This is a decreasing function of $\langle k \rangle$, for $\langle k \rangle_g < N$.

\subsection{Real data}
To test our algorithm on real data, we use the American College Football Network (ACFN) dataset provided in Ref.~\cite{girvan2002community}, which represents the schedule of Division I games for the season of the year 2000. Each node in the data set corresponds to a team, and each link is a game played between teams.
Teams are grouped in conferences, and each team plays most of its games within a same conference (though not all teams within a conference encounter each other). Conferences group teams of similar level, but another main criterion is geographic distance.
Therefore, this dataset has a community structure which is not based on affinity. 
Here, affinity indicates that teams are of similar level, and therefore should play in the same conference, if conferences were based solely on affinity. 

\begin{figure}[t!]
 \includegraphics[width=0.4\textwidth]{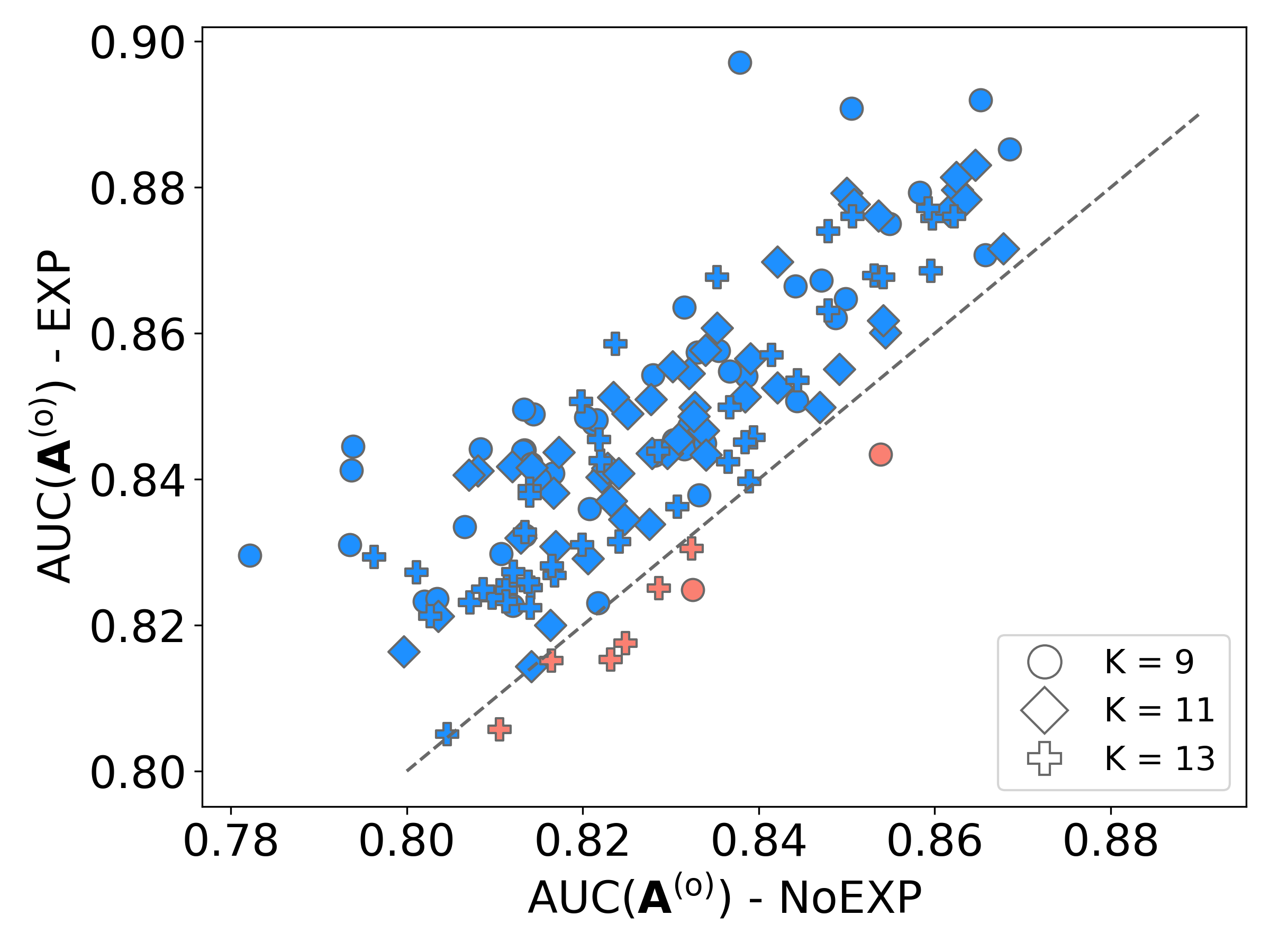}
 \includegraphics[width=.6\textwidth, trim=0 0 0 500]{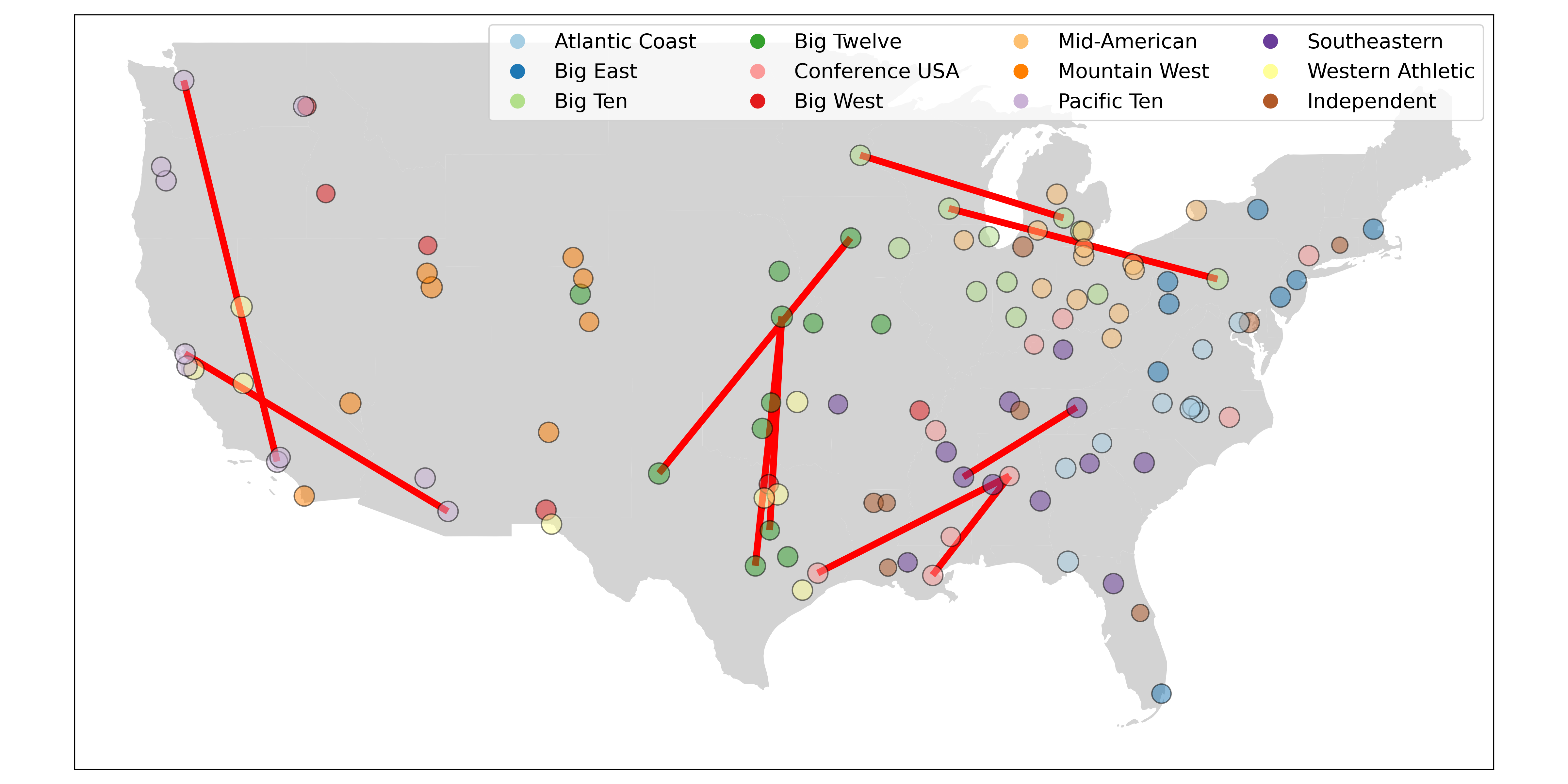}
 \caption{\textbf{Left}: Performance in predicting missing links of the \xp\ and \nxp\ algorithms on the ACFN dataset. Different marker shapes correspond to different numbers of communities, while blue (red) markers denote instances where \xp\ (\nxp) has better performance than \nxp\ (\xp). There is a total of 150 markers, denoting 5 folds repeated for 10 random seeds for each value of $K$. 
 \textbf{Right}: Top 10 games that are recommended by the \xp\ algorithm with $K=11$, which were not played in the AFCN data set. Different colors indicate different conferences.
 }
 \label{fig:real}
\end{figure}

We randomly hide 20\% of the links in the ACFN and check how well the \xp\ and \nxp\ algorithms are able to reconstruct which links are missing. We run the algorithm with various number of communities $K=9,11,13$, finding the best result at $K=11$, which is also the number of conferences in the dataset.
In \Cref{fig:real}--left we show a scatter plot of the AUC trial-by-trial. This reveals a superior performance of the \xp\ method which outperforms \nxp\ in 142 out of 150 trials (5 fold per 10 random seeds for each of $K=9,11,13$). This suggest that \xp\ is better capturing the data.

In \Cref{fig:real}--right we show the top 10 recommendations that we can extract from the \xp\ algorithm by taking, among the links missing from $\Ao$, those with smallest predicted exposure $Q_{ij}$ and the highest affinity $\lambda_{ij}$. Although, in the absence of ground truth, we are not able to assess the validity of these suggestions, we note that all the suggested links represent unplayed games within the the same conference
and that games within teams in different conferences were ranked lower.

\section{Conclusions}
In networks, nodes that would enjoy a high mutual affinity often appear disconnected for reasons that are independent of affinity. This is the case, for example, with people or entities in social networks that have never met, or due to some kind of sampling bias.
This introduces a sampling bias in the datasets used for community detection. We studied this problem through a  general framework, where we postulate that affinity in terms of compatibility of communities is not enough in order to explain the existence of a link, but rather a mechanism of exposure between nodes should be taken into account as well.

We proposed a principled probabilistic model, \xp, that takes into account this type of bias and is able to estimate the probability that two non-connected nodes are exposed while jointly learning what communities they belong to. We tested the \xp\ algorithm against a version of itself that does not account for exposure, \nxp. On artificial data, where we could validate our results on ground truth parameters and unobserved ground truth data, we found that \xp\ is as good as \nxp\ in learning communities, but it outperforms it when it comes to reconstructing missing links. In addition, the \xp\ approach allows us to satisfactorily infer which links remained unexposed, an estimate that cannot be done with standard method as, for example, \nxp. We finally tested our algorithm on a real dataset which has a hidden structure that is independent of the affinity between links, finding that also here the \xp\ algorithm is better at reconstructing missing links.

The principled approach that we used based on statistical inference is general. It can be made more specific depending on the application at hand. For example, we considered the simple case where exposure only depends on each individual's propensity towards being exposed. However, this could depend on a more fine structure of society, and we could think of introducing an exposure mechanism that mimics the presence of communities which are independent of affinity (\textit{e.g.} different schools, or different classes in a school). Allowing for community-dependent exposure has the potential to better mimic the kind of dilution that occurs in many real datasets. This can also apply to the AFCN dataset, where a better way to model exposure may be one that allows a structure that is able to account for different conferences or geographical regions. We leave this for future work. Additionally, exposure could be driven by covariate information on nodes, as also used in recommender systems \cite{liang2016modeling}. This could be integrated using variants of community detection methods that account for this extra information \cite{contisciani2020community,newman2016structure,fajardo2021node}.
Exposure could also change through time, and it could also have some dependence on the structure of $\Ag$. These are all interesting avenues for future work.

%
% ---- Bibliography ----
%
\bibliographystyle{unsrt}
\bibliography{bibliography}

\begin{thebibliography}{10}

\bibitem{fortunato2010community}
Santo Fortunato.
\newblock Community detection in graphs.
\newblock {\em Physics reports}, 486(3-5):75--174, 2010.

\bibitem{newman2018networks}
Mark Newman.
\newblock {\em Networks}.
\newblock Oxford university press, 2018.

\bibitem{fortunato2016community}
Santo Fortunato and Darko Hric.
\newblock Community detection in networks: A user guide.
\newblock {\em Physics reports}, 659:1--44, 2016.

\bibitem{liang2016modeling}
Dawen Liang, Laurent Charlin, James McInerney, and David~M Blei.
\newblock Modeling user exposure in recommendation.
\newblock In {\em Proceedings of the 25th International Conference on World
  Wide Web}, pages 951--961. International World Wide Web Conferences Steering
  Committee, 2016.

\bibitem{yang2018unbiased}
Longqi Yang, Yin Cui, Yuan Xuan, Chenyang Wang, Serge Belongie, and Deborah
  Estrin.
\newblock Unbiased offline recommender evaluation for missing-not-at-random
  implicit feedback.
\newblock In {\em Proceedings of the 12th ACM Conference on Recommender
  Systems}, pages 279--287, 2018.

\bibitem{wang2016learning}
Xuanhui Wang, Michael Bendersky, Donald Metzler, and Marc Najork.
\newblock Learning to rank with selection bias in personal search.
\newblock In {\em Proceedings of the 39th International ACM SIGIR conference on
  Research and Development in Information Retrieval}, pages 115--124, 2016.

\bibitem{chuklin2015click}
Aleksandr Chuklin, Ilya Markov, and Maarten~de Rijke.
\newblock Click models for web search.
\newblock {\em Synthesis lectures on information concepts, retrieval, and
  services}, 7(3):1--115, 2015.

\bibitem{de2017community}
Caterina De~Bacco, Eleanor~A Power, Daniel~B Larremore, and Cristopher Moore.
\newblock Community detection, link prediction, and layer interdependence in
  multilayer networks.
\newblock {\em Physical Review E}, 95(4):042317, 2017.

\bibitem{ball2011efficient}
Brian Ball, Brian Karrer, and Mark~EJ Newman.
\newblock Efficient and principled method for detecting communities in
  networks.
\newblock {\em Physical Review E}, 84(3):036103, 2011.

\bibitem{zhao2012consistency}
Yunpeng Zhao, Elizaveta Levina, and Ji~Zhu.
\newblock Consistency of community detection in networks under degree-corrected
  stochastic block models.
\newblock {\em The Annals of Statistics}, 40(4):2266--2292, 2012.

\bibitem{schein2016bayesian}
Aaron Schein, Mingyuan Zhou, David Blei, and Hanna Wallach.
\newblock Bayesian poisson tucker decomposition for learning the structure of
  international relations.
\newblock In {\em International Conference on Machine Learning}, pages
  2810--2819. PMLR, 2016.

\bibitem{safdari2021generative}
Hadiseh Safdari, Martina Contisciani, and Caterina De~Bacco.
\newblock Generative model for reciprocity and community detection in networks.
\newblock {\em Physical Review Research}, 3(2):023209, 2021.

\bibitem{contisciani2021community}
Martina Contisciani, Hadiseh Safdari, and Caterina De~Bacco.
\newblock Community detection and reciprocity in networks by jointly modeling
  pairs of edges.
\newblock {\em arXiv preprint arXiv:2112.10436}, 2021.

\bibitem{safdari2022reciprocity}
Hadiseh Safdari, Martina Contisciani, and Caterina De~Bacco.
\newblock Reciprocity, community detection, and link prediction in dynamic
  networks.
\newblock {\em Journal of Physics: Complexity}, 3(1):015010, 2022.

\bibitem{liben2007link}
David Liben-Nowell and Jon Kleinberg.
\newblock The link-prediction problem for social networks.
\newblock {\em Journal of the American society for information science and
  technology}, 58(7):1019--1031, 2007.

\bibitem{lu2011link}
Linyuan L{\"u} and Tao Zhou.
\newblock Link prediction in complex networks: A survey.
\newblock {\em Physica A: statistical mechanics and its applications},
  390(6):1150--1170, 2011.

\bibitem{girvan2002community}
Michelle Girvan and Mark~EJ Newman.
\newblock Community structure in social and biological networks.
\newblock {\em Proceedings of the national academy of sciences},
  99(12):7821--7826, 2002.

\bibitem{contisciani2020community}
Martina Contisciani, Eleanor~A Power, and Caterina De~Bacco.
\newblock Community detection with node attributes in multilayer networks.
\newblock {\em Scientific reports}, 10(1):1--16, 2020.

\bibitem{newman2016structure}
Mark~EJ Newman and Aaron Clauset.
\newblock Structure and inference in annotated networks.
\newblock {\em Nature communications}, 7:11863, 2016.

\bibitem{fajardo2021node}
Oscar Fajardo-Fontiveros, Roger Guimer{\`a}, and Marta Sales-Pardo.
\newblock Node metadata can produce predictability crossovers in network
  inference problems.
\newblock {\em Physical Review X}, 12(1):011010, 2022.

\end{thebibliography}

\appendix
\section{Asymmetric links} \label{app:asym}
Here, we show how to extend the \xp\ model to asymmetric networks ($A_{ij}\neq A_{ji}$), while still maintaining a symmetric structure for the exposure ($\mu_{ij}=\mu_{ji}=\mu_i\mu_j$). 

We introduce an extra set of $N \times K$-dimensional hidden variables, $v_{ik}$, which have the same structure of the $u_{ik}$. The latent variable set is now $\theta=(u,v,w)$. This allows us to replace Eq.~\eqref{eq:lambda} with an asymmetric Poisson parameter
\begin{equation}
    \lij=\sum_{k,q}^K u_{ik}v_{jq} w_{kq}\,.   
\end{equation}
As a result, Eqs.~\eqref{eqn:u}, \eqref{eqn:w}, \eqref{eqn:rho} and \eqref{eqn:mu} are replaced by
\begin{align}
u_{ik} &= \f{\sum_{j} Q_{ij}\, A_{ij}\sum_{q}\rho_{ijkq} }{\sum_{j}Q_{ij}\, \sum_{q}v_{jq}w_{kq}} \label{eqn:u-asym} \\
v_{iq} &= \f{\sum_{j} Q_{ij}\, A_{ij}\sum_{k}\rho_{ijkq} }{\sum_{j}Q_{ij}\, \sum_{k}v_{jq}w_{kq}} \label{eqn:v-asym} \\
w_{kq} & = \f{\sum_{i,j} Q_{ij}\, A_{ij}\rho_{ijkq} }{\sum_{i,j}Q_{ij}\, u_{ik}v_{jq}} \label{eqn:w-asym} \\
\rho_{ijkq} & =\f{u_{ik}v_{jq}w_{kq}}{\sum_{k,q}u_{ik}v_{jq}w_{kq}} \label{eqn:rho-asym} \\
\mu_{i} & =\f{\sum_{j}Q_{ij}} {\sum_{j}\f{(1-Q_{ij})\, \mu_{j} }{ (1-\mu_{i}\, \mu_{j})}} \label{eqn:mu-asym} \quad,
\end{align}
where now we also take into account the updates for the new hidden variables, $v_{ik}$.

\end{document}